\documentclass[epj]{svjour}
\usepackage{graphics}
\usepackage{amssymb}
\begin{document}
\title{Reliability of rank order in sampled networks}
%\author{First author\inst{1} \and Second author\inst{2}% etc
\author{P.-J. Kim \and H. Jeong\thanks{\email{hjeong@kaist.ac.kr}}
% \thanks{\emph{Present address:} Insert the address here if needed}%
}                     % Do not remove
\institute{Department of Physics, Korea Advanced Institute of Science and Technology, Daejeon 305-701, Korea}
\date{Received 26 May 2006 / Received in final form 29 November 2006}
\abstract{
In complex scale-free networks, ranking the individual nodes based upon 
their importance has useful applications, such as the identification of hubs for epidemic control, 
or bottlenecks for controlling
traffic congestion. However, in most real situations, only limited sub-structures 
of entire networks are
available, and therefore the reliability of the order relationships in sampled networks 
requires investigation.
With a set of randomly sampled nodes from the underlying original networks, 
we rank individual nodes by
three centrality measures: degree, betweenness, and closeness. 
The higher-ranking nodes from the sampled
networks provide a relatively better characterisation of their ranks in the original networks 
than the lower-ranking
nodes. A closeness-based order relationship is more reliable than any other quantity, due to the
global nature of the closeness measure. In addition, we show that if access to hubs is limited during the
sampling process, an increase in the sampling fraction can in fact decrease the sampling accuracy. Finally,
an estimation method for assessing sampling accuracy is suggested.
\PACS{
      {89.75.Hc}{Networks and genealogical trees}   \and
      {89.75.Fb} {Structures and organization in complex systems}
     } % end of PACS codes
} %end of abstract
\maketitle
\section{Introduction}
\label{intro}
In recent years, there has been great interest in examining
the properties of complex
networks such as the World Wide Web, the Internet, and social and biological networks \cite{Albert}.
Recent research on the networks reveals that
many networks have scale-free structures that possess a
right-skewed degree distribution. This power-law degree
distribution guarantees a noticeable existence of nodes, or
hubs, that have a very large number of connections compared
with the average node. The essential role of hubs in
networks is widely recognised in the contexts of immunisation
in epidemic spreading \cite{Vespignani}, the formation of social trends \cite{Luis},
finding drug targets on biological molecules \cite{Jeong,Jeong2}, and
optimal path finding strategy \cite{BJKim}.
For example,
the study of the spread of viruses on the Internet shows
that targeting immunisation on hubs drastically reduces
the occurrence of endemic states, even with a very low immunised
fraction, whereas uniform immunisation does not
lead to a drastic reduction in the infection prevalence \cite{Albert,Vespignani}.
For drug target identification in biological systems, the
likelihood that removal of a protein will be lethal correlates
strongly with the number of connections to that protein
in the protein-protein interaction network \cite{Jeong2}. 

In these examples, accurate identification of the important nodes, i.e.
hubs, is an efficient way to resolve
the specified problems.
Such identification, however, requires
that the ranks of the individual nodes are known, based on
their importance and contribution to the entire network.
In targeted immunisation on hubs, the more accurate a
policy is at identifying the ranks, the smaller the number
of necessary cures \cite{Dezso}.
In real situations, however, only
part of the information on the underlying networks can
be exploited, due to severe physical and economical constraints \cite{BJKim,Frank,Amaral,Koss,Scholz,Koss2}.
For example, a survey of relationships
among participants has to be conducted in order to construct
a social network, but the collected network data
might be incomplete, since surveys usually target only a
limited sample of the whole population. Therefore, the
statistical properties of a network must frequently be assessed
without complete knowledge of global information
on the entire network. Nevertheless, the sampling problem
in complex networks has not yet been extensively explored
\cite{Koss,Koss2,Kim}, despite the substantial interest in the
community of social network analysis \cite{Frank}.

Given that only partial information on a network can
be obtained, it is worth investigating how accurately the
importance of a node, based on only partial information,
reflects the actual importance of the node in the original
network. For successful epidemic control, it is important
to determine whether or not the hubs identified as critical
from the incomplete data remain so even after adding supplementary
data \cite{aids}. The study of rank reliability in sampled networks can also be applied to
many technological and biological systems,
and avoids possible artifacts depending on a specific numerical scale of data (whereas stretching or compressing the scale does not alter a rank-based result).

In the present work, we analyse the Barab\'asi-Albert (BA) model as the prototype example of a scale-free network 
\cite{Albert}~\footnote{Performing the analysis on the configuration model \cite{random} 
instead of the BA model does not alter the current results.},
which allows us to clearly discriminate the contribution
of the power-law degree distribution to the sampling
effect from the contribution of additional specific
biases that appear in other networks. Furthermore, to
consider realistic effects that are disregarded in the BA
model, we also analyse several real networks, such as the
Los Alamos e-Print Archive coauthorship network
\cite{Newman}, the Internet AS \cite{DIMES}, and protein-protein interaction networks \cite{Jeong2,Goh2}.
We concentrate only on cases where the accessible
information on the networks is limited to the connectivity
between randomly sampled nodes, although in
reality, there are other kinds of allowable information, including
the connectivity from snowball sampling, and that
from randomly sampled links \cite{Kim}.
Snowball samples consist
of identified nodes to which all linked nodes are then
used to refer to other nodes, and are usually employed by
Web search engines. Randomly sampled links describe the
randomly gathered connectivity between nodes, e.g. in the
case of poorly gathered contact information between patients.
It is expected that snowball sampling provides rare
sampling biases with literally conserved topologies during
the sampling, while the possible nontrivial results from
randomly sampled links can be sufficiently analogous to
those from randomly sampled nodes with some correspondence
between them \cite{Kim}.
Thus, the focus on randomly
sampled nodes could be considered a reasonable step towards
investigating the network-sampling problems, although
the study of only randomly sampled nodes here
has limitations for understanding more specific problems
in real situations. In this regard, the possible deviations
between our results and reality could be further reduced
by additional investigation of different sampling schemes.

\section{Measured quantities}
\label{mesure}
In sampled or entire networks, individual nodes can be properly ranked
according to their importance or {\it prestige} \cite{prst}, like degree.
With the set of sampled nodes, we first define a measure for the rank correlation
between the sampled nodes and the nodes in the original network,
denoted by $\tau$, which is a variant of {\it Kendall's~Tau} \cite{Johnson},
representing how faithfully the rank order is preserved in the sampled network.
For an arbitrary pair of sampled nodes $\{i, j\}$,
the nodes have the assigned importance, like degrees, such as $\{k_{i},k_{j}\}$
in the sampled network and as $\{k^{o}_{i},k^{o}_{j}\}$ in the original network.
If $k_{i}<k_{j}$ ($k_{i}>k_{j}$) and $k^{o}_{i}<k^{o}_{j}$ ($k^{o}_{i}>k^{o}_{j}$),
or $k_{i}=k_{j}$ and $k^{o}_{i}=k^{o}_{j}$, we consider that the pair is then ordered similarly
in the sampled and original networks. Otherwise, it is regarded as ordered dissimilarly.
To quantify the preservability of rank order, the dominance of pairs ordered similarly
in both the sampled and original networks is normalised
by the total number of pairs that are considered in the calculation,
through $\tau = (N_{+}-N_{-})/(N_{+}+N_{-})$, where $N_{+}$ is the number of
pairs ordered similarly for sampled and original networks, and
$N_{-}$ is the number of pairs that are ordered dissimilarly.
$\tau$ can have a value from $-1$ to $1$, indicating complete disagreement and full agreement, respectively.
Without any tied ranks, if the ranks are more preserved in sampling than expected by random shuffling, $\tau$ is positive.
For the probability $p$ that an arbitrary pair is ordered similarly, we can obtain the relationship $p = (\tau +1)/2$.

Because the statistical properties of many real networks
follow a universal characteristic like a power-law
distribution, their preservability in sampled networks has
been of basic interest in previous studies \cite{Koss,Kim}.
These statistical properties, however, are rarely affected by interchanging
the prestiges of nodes. Hence, it is worth
comparing the preservability of these individual-prestige--insensitive properties 
in sampled networks to that of the
individual-prestige--sensitive properties such as $\tau$.
Therefore, we introduce another complementary measure, $\rho$, which represents the similarity between two probability distributions of centrality -- 
one from sampled nodes and the other from the original network -- where the latter one obeys a power law. First,
we obtain the cumulative distribution of variable $k$, $\mathcal{P}_{S}(k)$ from the sampled nodes, and $\mathcal{P}_{O}(k)$ from the original network.
Using $k_{i}$ of the $i$th sampled node, we find $k^{o}_{i}$ satisfying that $\mathcal{P}_{S}(k_{i})=\mathcal{P}_{O}(k^{o}_{i})$, 
and calculate the Pearson correlation $\rho$ between $k_{i}$ and $k^{o}_{i}$ for $i=1,2,\dots,N$ where $N$ is the number of sampled nodes.
$\rho$ can achieve its maximum value $1$
if $k_{i}$ is proportional to $k^{o}_{i}$.
This means that when $\mathcal{P}_{S}(k)\varpropto{k}^{-\alpha}$ and $\mathcal{P}_{O}(k)\varpropto{k}^{-\beta}$, $\rho$
can achieve its maximum value $1$ if ${\alpha}={\beta}$, i.e. in
the case of identical power-law distributions. By applying
proper normalisation, we transform the measure $\rho$ so as to
take a value from $0$ to $1$ in its significant range~\footnote{To this end, we calculate the Pearson correlation $\rho_{th}$ between $k_{i}$ and $k^{o}_{i}$ as if $\mathcal{P}_{S}(k)$ is a simple linear function of $k$. 
We finally obtain the value of $\rho$ as ${\rho}\rightarrow (\rho-\rho_{th})/(1-\rho_{th})$. Therefore, $\rho$ becomes positive 
if the probability distribution from the sampled nodes resembles that from the original network more than a simple linear curve does.}.
$\rho$ gives the preservability of probability distributions rather than that of the node rank, thus $\rho$ can have a large value 
under the similar probability distributions, even if the ranks themselves are severely altered.
In a practical sense, it is possible to directly evaluate the
exponent difference of the power-law distribution between
sampled and original networks, and the detail of the results
exhibits some notable properties, including a slight
overestimation of the exponents during the random sampling \cite{Kim}. 
For the degree distribution of the BA model, a sampling overestimation of the exponent by a factor of 1.2
corresponds to $\rho=0.8$.
It should be noted that
an isolated node, which had no links to the other connected sampled nodes,
was excluded in the calculation of $\tau$ and $\rho$.
We can easily apply these measures to other quantities, such as
betweenness centrality, as will be shown below.

\section{Simulation and results}
\label{result}

\begin{figure}[h]
\begin{center}
\resizebox{0.95\columnwidth}{!}{
\includegraphics{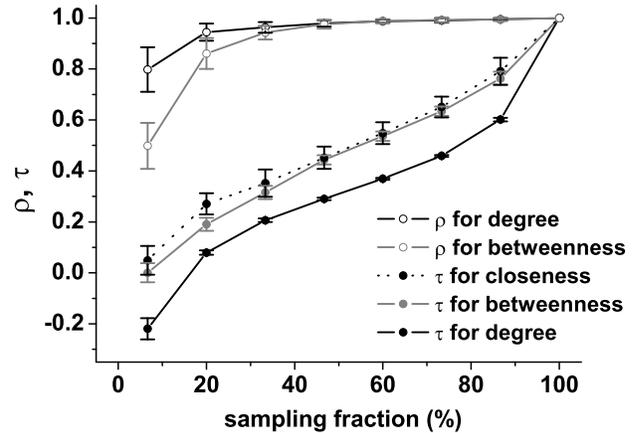}
}
\caption{The horizontal axis represents the sampling fraction (the number of sampled nodes divided by the number of total nodes),
while the vertical axis represents the calculated $\tau$ and $\rho$ at each sampling fraction. We use the BA model with 30000 nodes and average degree of 8.}
\label{BAall}
\end{center}
\end{figure}

In this paper, we rank-order individual nodes using the three centrality measures of complex networks;
degree, betweenness, and closeness, in order to calculate $\tau$ \cite{Newman,Freeman}~\footnote{
Closeness of the $i$th node is defined as the average of the reciprocal distances from
 the $i$th node to all other nodes. 
}.
We also calculate $\rho$ for degree and betweenness, based on their power-law statistics \cite{Albert,Goh}.
Using randomly sampled nodes, Figure~\ref{BAall} displays the result for the
BA model, which reflects results typical for other real networks
with regard to the qualitative distinction between $\tau$ and $\rho$.
In Figure~\ref{BAall}, as the sampling fraction increases,
$\tau$ grows gradually while $\rho$ grows quickly and saturates at $1$~\footnote{
 Randomly sampled nodes are inevitably composed of several disconnected clusters.
 Nonetheless, the main difference between $\tau$ and $\rho$ is still valid even if
 we consider only the largest component from these clusters.
}.
It has been verified that the early saturation of $\rho$ is due to the overall proportional relation
between the centrality measure obtained from the randomly sampled nodes and that from the original networks \cite{Kim}.
On the other hand, the continuous and rather slow growth of $\tau$ indicates
the sensitivity of individual-level prestige to the sampling,
especially for the low rank nodes, as will be presented below.

\begin{figure}[h]
\begin{center}
\resizebox{0.95\columnwidth}{!}{
\includegraphics{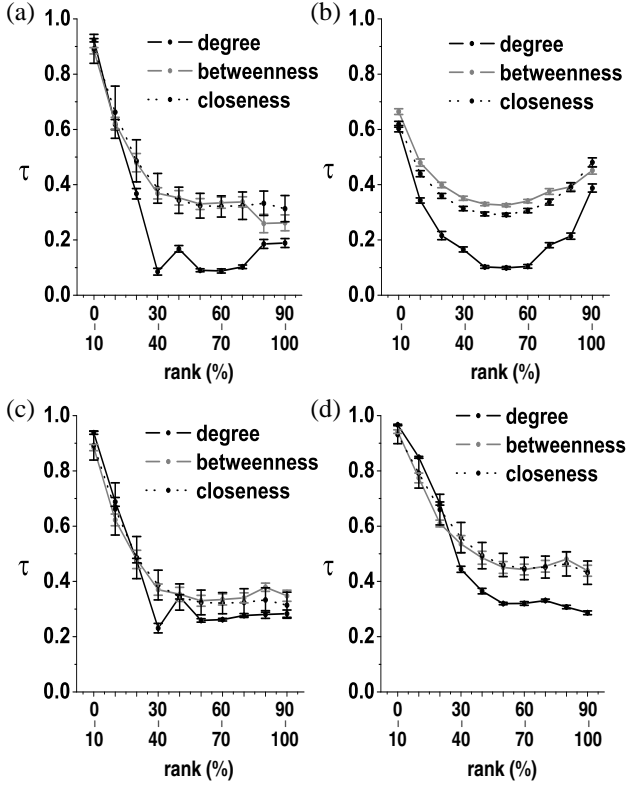}
}
\caption{The horizontal axis represents the groups according to their constituting ranks
in sampled nodes, while
the vertical axis represents $\tau$ for each group.
(a) $40\%$ sampled from the BA model with 30000 nodes and an average degree of 8;
(b) the same condition as (a) in the Erd\"os-R\'enyi model;
(c) $\tau$ calculated without counting the pairs in tied prestiges from (a);
(d) increased sampling fraction with $60\%$ from (c).}
\label{Eqi}
\end{center}
\end{figure}

Interestingly, the contribution of an individual node to the value of $\tau$ is not uniform over all nodes,
and strongly depends on the rank of the node.
To examine this property in detail,
we divide the sampled nodes into the subgroups according to their individual ranks
in the sampled nodes.
For example, in the case of degree-based ranks, each node would belong to one of 10 groups
-- the highest $0{\sim}10\%,10{\sim}20\%,\dots,90{\sim}100\%$ ranks --
in descending order of degree.
To obtain the contribution to $\tau$ made by each group, we calculate $\tau$ over pairs of nodes $\{i, j\}$
where the $i$th node is a member of the given group, and the $j$th node is a member of any group. Figure~\ref{Eqi}a illustrates the result for the BA model;
the groups of higher-ranking nodes have large $\tau$'s, indicating that the higher-ranking nodes 
of the sampled nodes provide better characterisation of their ranks in the original networks \cite{Scholz}.
This point will be universally carried in scale-free networks,
because the nodes of large degree hardly face the shuffling of their ranks
in sampling due to their relatively small population.
In the Erd\"os-R\'enyi model, the intermediate ranks comprise a greater proportion of the population than either the high or low ranks. It is expected that $\tau$ would reach the minimum value with intermediate ranks, as observed in Figure~\ref{Eqi}b.

From observations in the BA model, we have found that
$\tau$'s for betweenness and closeness have larger values than $\tau$ for degree,
except in a few of the highest rank groups.
Even in these highest rank groups, $\tau$ for degree is comparable to the other $\tau$'s, and does not 
dominate them. In an attempt to explain the smallness of $\tau$ for degree in most groups, 
one might consider the discreteness effect of degree, e.g. the majority of the nodes 
would possess a degree of $1$ in a small sampling fraction.
This severe discreteness could hide the original ordinal information between the nodes, 
thus leading to a smaller value of $\tau$.
Nevertheless, the discreteness effect does not sufficiently explain our observation.
To clarify this point, we calculate $\tau$ while excluding the pairs of sampled nodes in tied prestiges, 
which reduces the discreteness effect.
Figures~\ref{Eqi}c and \ref{Eqi}d show the results of sampling fractions of $40\%$
and $60\%$, respectively, but well represent the generic consequence
along all sampling fractions.
Although $\tau$ for degree becomes large in a small sampling fraction (see Fig.~\ref{Eqi}c), 
the similar feature in Figure~\ref{Eqi}a eventually recovers
as the sampling fraction increases (see Fig.~\ref{Eqi}d).
This result implies that the small value of $\tau$ for degree can be
attributable to the intrinsic property
of the local centrality, by which individual prestige is sensitive to the random sampling
due to the local fluctuation of the network topology.

\begin{figure}[t]
\begin{center}
\resizebox{0.95\columnwidth}{!}{
\includegraphics{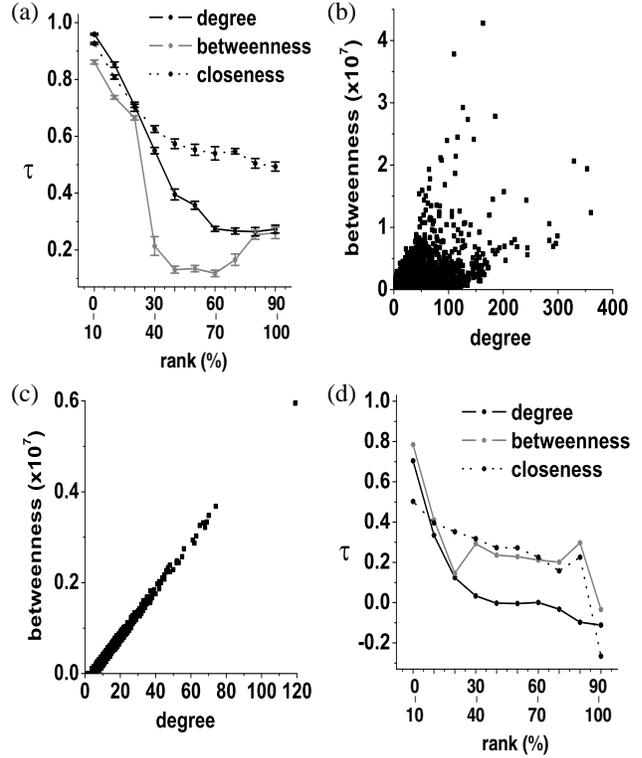}
}
\caption{
(a) Rank vs. $\tau$ in the $30\%$ sampled nodes from the Archive coauthorship.
(b) Individual degree vs. betweenness in the Archive coauthorship.
(c) Individual degree vs. betweenness in the highest $30\%$ of highly correlated degree and
betweenness.
(d) Rank vs. $\tau$ of the nodes from (c).
}
\label{bcstg}
\end{center}
\end{figure}

For comparison with the BA model, we consider real networks, and observe some different results.
In real networks, $\tau$ for betweenness becomes suppressed and is no longer comparable to $\tau$ 
for closeness (see Fig.~\ref{bcstg}a).
Here, we present the case of
the Los Alamos e-Print Archive coauthorship network, although similar results are observed 
in other real networks,
including the Internet AS and protein-protein interaction networks.

The suppressed $\tau$ for betweenness reflects the sensitivity
of the betweenness measure to the network modularity.
Unlike random networks including the BA model,
many real networks have structural sub-units, namely modular structures, that significantly affect
the centrality measures in unexpected ways in random networks.
For example, the presence of nodes with small degree and large betweenness
shown in Figure~\ref{bcstg}b indicates the existence of loose connections between
tightly-knit modules \cite{Girvan}, such that
the nodes on the loose connections that bear a considerable number of inter-modular communication paths
exhibit large betweenness centrality despite their small degree.
In this sense, during the random sampling, violating
modularity in the networks can significantly alter the
betweenness-based node prestige, thereby lowering $\tau$ for betweenness.
One way to confirm this effect is to observe what happens if the modularity effect is reduced.
To discard the modularity effect,
we sample only the nodes with highly correlated degree and betweenness rather than do random sampling, 
and calculate the corresponding $\tau$~\footnote{
 We choose the nodes in ascending order of
 \begin{displaymath}
 \Bigg\vert \frac{x_{i}-\langle x_{i}\rangle}{\sqrt{\langle x_{i}^{2}\rangle -\langle x_{i}\rangle^{2}}}
 -\frac{y_{i}-\langle y_{i}\rangle}{\sqrt{\langle y_{i}^{2}\rangle -\langle y_{i}\rangle^{2}}} \Bigg\vert\,\,,
 \end{displaymath}
 where $x_{i}$, $y_{i}$ stand for the degree and betweenness of
 the $i$th node, and $\langle \cdots\rangle$ is the average over all nodes.
}.
Under the reduced-modularity effect, we can identify the range of the sampling fraction 
(in the Archive coauthorship, ${\lesssim}50\%$) in which $\tau$ for betweenness becomes comparable 
to $\tau$ for closeness as in the BA model (see Figs.~\ref{bcstg}c and \ref{bcstg}d). 
This shows that the modularity effect is indeed essential to the suppression of $\tau$ for betweenness.

Consequently, the $\tau$ for each centrality measure relies on the sensitivity of the centrality measure 
to the sampling.
Indeed, a small $\tau$ for degree comes from the fact that the ranks
of degree suffer from their shuffling due to the local fluctuation of topology during the sampling process.
Although it is based upon global information on the networks, betweenness concerns the number 
of {\it shortest} paths across a node itself, thus the rank can be sensitive to the topological variation 
in the proximity of the node, and especially to modular-level fluctuations.
On the other hand, closeness is relatively tolerant to such topological fluctuations, 
and contributed to by the robust global information of the network, averaged path lengths outward from a node.
Therefore, the closeness-based rank order possesses a larger $\tau$ than any other quantity, 
due to the unique globality of the closeness being insensitive to the sampling.

\begin{figure}[h!]
\begin{center}
\resizebox{0.95\columnwidth}{!}{
\includegraphics{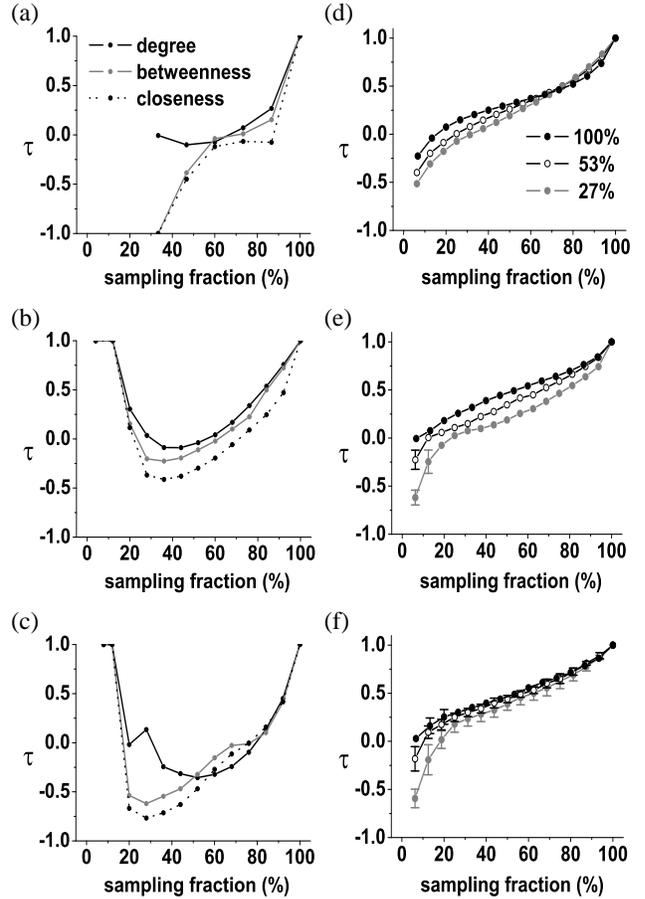}
}
\caption{(a)--(c) Sampling under limited hub-accessibility.
The horizontal axis represents the sampling fraction in ascending order of degree,
while the vertical axis represents $\tau$ at each sampling fraction, for
(a) the BA model with 30000 nodes and average degree of 8;
(b) the Archive coauthorship; and
(c) a random network given the same degree distribution as that of the Archive coauthorship.
Results similar to those shown in (b) and (c) are also shown for other real networks.
(d)--(f) Sampling from randomly sampled networks.
The horizontal axis for a sampling fraction out of $100\%$-, $53\%$-, and $27\%$-sampled nodes
from the BA model with 30000 nodes and average degree of 8, and
the vertical axis for $\tau$ at each sampling fraction.
(d) $\tau$ for degree.
(e) $\tau$ for betweenness.
(f) $\tau$ for closeness.
}
\label{smplin}
\end{center}
\end{figure}

Because such a global characteristic of closeness is responsible for the large $\tau$ for closeness,
the value of $\tau$ for closeness can be suppressed if
access to the hubs that bind the network together globally is restricted in the sampling process.
To simplify this situation, we sample the nodes in ascending order of 
their centrality measures, rather than randomly as presented before.
Figures~\ref{smplin}a--\ref{smplin}c display the results gathered 
when nodes are selected in ascending order of degree,
and similar results are produced for the cases of betweenness and closeness.
As discussed above, the value of $\tau$ for closeness is no longer superior to
any other quantity.
Surprisingly, we further identify that in real networks, $\tau$ obtains its minimum value
in an intermediate range of the sampling fraction, and thus has a convex shape (see Fig.~\ref{smplin}b).
This directly indicates that with small sampling fractions, if access to hubs is limited,
an increase in the sampling fraction (i.e., more nodes are sampled)
can in fact decrease the sampling accuracy (small $\tau$)
without a gain in valuable information.
To avoid this type of error in the analysis of social networks,
a sufficient sampling size of social individuals must be assured when access to the central leadership is restricted.
Also, for the study of small data sets in bioinformatics, the presence of hubs should be of concern
because if they are not available,
the ordinal information extracted from the small data set is not reliable.
This exotic behaviour from real networks is essentially caused by
the properties of the degree distribution of real networks
rather than by other structural properties embedded in real networks, e.g. the modularity.
Figure~\ref{smplin}c exhibits the results for the random networks
given the same degree distribution as that of the real networks \cite{random},
which produce a feature similar to that shown in Figure~\ref{smplin}b.

For predictive purposes, is it possible to presume the value of $\tau$ for nodes 
sampled randomly from entire networks?
In real situations, since the information available to us is that of sampled networks
rather than that of entire networks, we can only evaluate the $\tau$ of the nodes
against {\it a priori} sampled networks, but not against the entire networks, which are rarely achievable.
Despite such limitations, we can exploit
the $\tau$ of the nodes sampled from these sampled networks
to approximate that from the entire networks.
In random sampling, since decreasing the sampling fraction makes the
network more homogeneous, with small degrees \cite{Kim}, 
it is expected that the $\tau$ of the subset in randomly sampled nodes underestimates
that of the subset in the entire network with the same sampling fraction.
For the BA model, Figures~\ref{smplin}d--\ref{smplin}f establish the corresponding tendency manifested especially
in low sampling fractions, which is consistently revealed in the cases of other real networks only except
for the betweenness of the Internet AS~\footnote{
 For the betweenness of the Internet AS, the tendency becomes reversed such that
 $\tau$ of the subset in randomly sampled nodes overestimates
 that of the subset in the entire network with the same sampling fraction.
}.
In this regard, we can use this underestimation to approximate the actual $\tau$
of an arbitrary sampling fraction for the entire network by providing its lower bound.
For example, in the case of the protein-protein interaction network,
$\tau$ for the degree of $30\%$ sampled nodes in our data is equal to $0.35$, which 
means that the $\tau$ for $30\%$ sampled nodes in a complete data set would be greater than $0.35$
if the sampling method is close to random node sampling.
Likewise, for $30\%$ sampled nodes in the Archive coauthorship,
$\tau$ for degree is equal to $0.45$, thereby indicating that $\tau$ would be larger than $0.45$
for the $30\%$ sampled nodes in the complete data set.

\section{Conclusions}
\label{conc}
In summary, we have investigated the accuracy of order relationships
in sampled networks, and found that the properties
of complex networks, such as degree heterogeneity
and structural modularity, are responsible for the various
results. The higher-ranking nodes in sampled networks
preserve their positions in the original networks more robustly
than the lower-ranking nodes, and the closeness-based
order relationship gives the best measure for faithful
ordinal information in sampled networks. Interestingly,
we discovered that limiting the access to hubs during the
sampling can in fact decrease the accuracy of the sampling
as the sampling fraction increases. We emphasise
the role of hubs in characterising a sampled network, and
the effect of the perturbed scale of the network, to which
each centrality measure responds sensitively. Beyond these
analyses, a methodology providing the lower bound for
sampling accuracy is suggested. Our results can be helpful
for understanding the properties of sampled networks,
especially for social and criminal networks, for which analysis
suffers from various types of sampling error and other
limitations \cite{Frank,Koss2,Krebs}.
The sampling problems in complex
networks, including the detection of errors in power-law
statistics and the suggestion of useful sampling protocols,
are currently being explored \cite{Koss,Koss2,Kim}.

\begin{acknowledgement}
We thank Sang Hoon Lee, Kwang-Il Goh, Dong-Hee Kim, Yong-Yeol Ahn, and Seung-Woo Son for
providing valuable information and technological support.
P.K. acknowledges the support by KOSEF-ABRL Program through Grant No. R14-2002-059-01002-0,
and H.J. acknowledges support by the Ministry of Science and Technology, through Korean Systems Biology Research
Grant (M10309020000-03B5002-00000).
\end{acknowledgement}

%\subsection{Subsection title}
%\label{sec:2}
%as required. Don't forget to give each section
%and subsection a unique label (see Sect.~\ref{sec:1}).
%\begin{figure}
%\resizebox{0.75\columnwidth}{!}{%
% \includegraphics{leer.eps}
%\vspace{5cm}       % Give the correct figure height in cm
%\end{figure}
%
%
% BibTeX users please use
% \bibliographystyle{}
% \bibliography{}

\begin{thebibliography}{}
\bibitem{Albert}
 S.H. Strogatz, Nature {\bf 410}, 268 (2001);
 R. Albert, A.-L. Barab\'asi, Rev. Mod. Phys. {\bf 74}, 47 (2002);
 S.N. Dorogovtsev, J.F.F. Mendes, Adv. Phys. {\bf 51}, 1079 (2002);
 M.E.J. Newman, SIAM Rev. {\bf 45}, 167 (2003)
\bibitem{Vespignani}
 R. Pastor-Satorras, A. Vespignani, Phys. Rev. Lett. {\bf 86}, 3200 (2001);
 Phys. Rev. E {\bf 65}, 036104 (2002)
\bibitem{Luis}
 L.M.A. Bettencourt, cond-mat/0304321
\bibitem{Jeong}
 H. Jeong, B. Tomber, R. Albert, Z.N. Oltvai, A.-L. Barab\'asi, Nature {\bf 407}, 651 (2000)
\bibitem{Jeong2}
 H. Jeong, S.P. Mason, A.-L. Barab\'asi, Z.N. Oltvai, Nature {\bf 411}, 41 (2001);
 H. Jeong, Z.N. Oltvai, A.-L. Barab\'asi, ComPlexUs {\bf 1}, 19 (2003)
\bibitem{BJKim}
 B.J. Kim, C.N. Yoon, S.K. Han, H. Jeong, Phys. Rev. E {\bf 65}, 027103 (2002)
\bibitem{Dezso}
 Z. Dezs\"o, A.-L. Barab\'asi, Phys. Rev. E {\bf 65}, 055103 (2002)
\bibitem{Frank}
 O. Frank, {\it Models and Methods in Social Network Analysis}, edited by P.J. Carrington et al. (Cambridge University Press, New York, 2005), p31
\bibitem{Amaral}
 S. Mossa, M. Barth\'el\'emy, H.E. Stanley, L.A. Nunes Amaral, Phys. Rev. Lett. {\bf 88}, 138701 (2002)
\bibitem{Koss}
 A. Clauset, C. Moore, Phys. Rev. Lett. {\bf 94}, 018701 (2005);
 L. Dall\'Asta, I. Alvarez-Hamelin, A. Barrat, A. Vazquez, A. Vespignani, Theor. Comput. Sci. {\bf 355}, 6 (2006);
 J.-L. Guillaumea, M. Latapya, D. Magoni, Comput. Networks {\bf 50}, 3197 (2006); 
 M.P.H. Stumpf, C. Wiuf, R.M. May, Proc. Natl. Acad. Sci. U.S.A. {\bf 102}, 4221 (2005);
 J.-D.J. Han, D. Dupuy, N. Bertin, M.E. Cusick, M. Vidal, Nat. Biotechnol. {\bf 23}, 839 (2005)
\bibitem{Scholz}
 J. Scholz, M. Dejori, M. Stetter, M. Greiner, Physica A {\bf 350}, 622 (2005)
\bibitem{Koss2}
 E. Costenbader, T.W. Valente, Soc. Networks {\bf 25}, 283 (2003);
 G. Kossinets, cond-mat/0306335
\bibitem{Kim}
 S.H. Lee, P.-J. Kim, H. Jeong, Phys. Rev. E {\bf 73}, 016102 (2006)
\bibitem{aids}
 A.S. Klovdahl, Soc. Sci. Med. {\bf 21}, 1203 (1985)
\bibitem{Newman}
 M.E.J. Newman, Phys. Rev. E {\bf 64}, 016131 (2001); Phys. Rev. E {\bf 64}, 016132 (2001)
\bibitem{DIMES}
 Y. Shavitt, E. Shir, cs.NI/0506099
\bibitem{Goh2}
 K.-I. Goh, B. Kahng, D. Kim, J. Korean Phys. Soc. {\bf 46}, 551 (2005)
\bibitem{prst}
 S. Fortunato, A. Flammini, F. Menczer, cond-mat/0602081
\bibitem{Johnson}
 G.K. Bhattacharyya, R.A. Johnson, {\it Statistical~Concepts~and~Methods} (Wiley, New York, 1977)
\bibitem{Freeman}
 L.C. Freeman, Soc. Networks {\bf 1}, 215 (1979)
\bibitem{Goh}
 K.-I. Goh, B. Kahng, D. Kim, Phys. Rev. Lett. {\bf 87}, 278701 (2001)
\bibitem{Girvan}
 M. Girvan, M.E.J. Newman, Proc. Natl. Acad. Sci. USA {\bf 99}, 7821 (2002)
\bibitem{random}
 M.E.J. Newman, S.H. Strogatz, D.J. Watts, Phys. Rev. E {\bf 64}, 026118 (2001);
 M. Molloy, B. Reed, Random Struct. Algorithms {\bf 6}, 161 (1995);
 Combinatorics, Probab. Comput. {\bf 7}, 295 (1998)
\bibitem{Krebs}
 V.E. Krebs, Connections {\bf 24}, 43 (2002)

%Author, \textit{Book title} (Publisher, place year) page numbers
% etc
\end{thebibliography}

\end{document}